\documentclass{svproc}
\usepackage{graphicx}%
\usepackage{multirow}%
\usepackage{amsmath,amssymb,amsfonts}%
\usepackage{mathrsfs}%
\usepackage[title]{appendix}%
\usepackage{xcolor}%
\usepackage{textcomp}%
\usepackage{manyfoot}%
\usepackage{booktabs}%
\usepackage{algorithm}%
\usepackage{algorithmicx}%
\usepackage{algpseudocode}%
\usepackage{listings}%
\usepackage{tabularx}
\usepackage{xltabular}
\usepackage{orcidlink}
\usepackage{enumitem}
\usepackage[usestackEOL]{stackengine}
\usepackage{amsmath,amssymb,amsfonts}
\usepackage{appendix}
\usepackage{cite}
\usepackage{url}

\begin{document}
\mainmatter              
\title{Prestigious but less interdisciplinary: a network analysis on top-rated journals in medicine}
\titlerunning{Running title}  
%
\author{Anbang Du\inst{1}  Michael Head\inst{2}
Markus Brede\inst{1}}
\authorrunning{Anbang Du et al.} 

\institute{School of Electronic and Computer Science, University of Southampton, Southampton, UK\\
\email{ad1u21@soton.ac.uk},\\
\and
Faculty of Medicine, University of Southampton, Southampton, UK}

\maketitle              

\begin{abstract}
Interdisciplinary research, a process of knowledge integration, is vital for scientific advancements. It remains unclear whether prestigious journals that are highly impactful lead in disseminating interdisciplinary knowledge. In this paper, by constructing topic-level correlation networks based on publications, we evaluated the interdisciplinarity of more and less prestigious journals in medicine. We found research from prestigious medical journals tends to be less interdisciplinary than research from other medical journals. We also established that cancer-related research is the main driver of interdisciplinarity in medical science. Our results indicate a weak tendency for differences in topic correlations between more and less prestigious journals to be co-located. Accordingly, we identified that interdisciplinarity in prestigious journals mainly differs from interdisciplinarity in other journals in areas such as infections, nervous system diseases and cancer. Overall, our results suggest that interdisciplinarity in science could benefit from prestigious journals easing rigid disciplinary boundaries.


\end{abstract}
\section{Introduction}\label{sec1}
Science can be seen as a complex, self-organizing, and evolving network of scholars, projects, publications, and ideas \cite{fortunato2018science}. Scientific activities like knowledge production, dissemination, and collaboration have been widely studied through the lense of complex networks \cite{zengScienceSciencePerspective2017,fortunato2018science,duIntegrationVsSegregation2025a,Du2025}. Interdisciplinary research, defined as a process of knowledge integration \cite{rafolsDiversityNetworkCoherence2010,duIntegrationVsSegregation2025a}, is vital to scientific innovation and has been widely advocated by policymakers \cite{wagnerApproachesUnderstandingMeasuring2011a}. Recent research has shown that interdisciplinarity can be captured and measured in terms of weighted complex networks \cite{rafolsDiversityNetworkCoherence2010,duIntegrationVsSegregation2025a}.

The relationship between interdisciplinarity and impact has long been a topic of discussion. Some literature has suggested that interdisciplinary research tends to be more frequently cited \cite{chenAreTopcitedPapers2015,okamuraInterdisciplinarityRevisitedEvidence2019a}. In contrast, some other studies argue that this is not necessarily the case as the way how interdisciplinarity and citations are measured could influence the results \cite{molas-gallartRelationshipInterdisciplinarityImpact2014,lariviereRelationshipInterdisciplinarityScientific2010,yegros-yegrosDoesInterdisciplinaryResearch2015a}. A very recent paper has revealed a more nuanced relationship between citation and interdisciplinarity: papers with higher interdisciplinarity are more likely to experience delayed recognition and greater citation sustainability \cite{caiRelationshipInterdisciplinarityCitation2023}. Beyond scholarly impact, interdisciplinary research has also been found to attract more policy attention \cite{huInterdisciplinaryResearchAttracts2024b}.

Prestigious journals, or high-ranking journals, are known for their high visibility and strong influence over the ecosystem of science \cite{milojevicNatureSciencePNAS2020}. Yet, limited attention has been paid to assess how prestigious journals lead in disseminating interdisciplinary knowledge in science. In this paper, we will address this gap by modelling and comparing the structure of interdisciplinary knowledge published in prestigious and less prestigious medical journals as two temporal networks. We aim at answering two research questions: (a) whether prestigious journals produce more interdisciplinary knowledge than less prestigious journals in medical science? (b) can we characterize areas of research in which interdisciplinary research published in prestigious and less prestigious journals differs?


\section{Method}
In this paper, we are interested in comparing knowledge structures in medical publications published in higher (more prestigious) and lower impact (less prestigious) journals. Accordingly, we collected data from the PubMed database in 1999, 2010 and 2022. PubMed was chosen due to its comprehensive collection of published works in the biomedical sciences. 

Distinguishing lower and higher impact also requires a definition of a journal's impact. Traditional journal ranking like the Journal Citation Report (JCR) from the Web of Science reports field-specific quantile ranking, where the top 25\% (quantile 1) is generally deemed as high-impact journals in the field.\footnote{\url{https://support.clarivate.com/ScientificandAcademicResearch/s/article/Journal-Citation-Reports-Quartile-rankings-and-other-metrics?language=en_US}} In this paper, we defined a journal to be impactful in a year if it is one of the top 10\% journals in the SCImago Journal Rank (SJR) in medicine of that year, where the SJR is a widely used metric that weighs the value of a citation based on the subject field, quality and reputation of the source.\footnote{\url{https://www.elsevier.com/en-gb/products/scopus/metrics}}. 

We note that the ranking of journal impact has been subject to criticism, a major reason being the field-dependency of journal impact factors \cite{lariviereSimpleProposalPublication2016}. However, the present study only concerns journals in the field of medicine, so that field-specific differences in citation habits are unlikely to have a major impact on results. A detailed investigation of citation differences in sub-fields of medicine might deserve attention in the future.

Each article in the PubMed database is indexed with a list of Medical Subject Headings (MeSH),\footnote{MeSH is a curated hierarchically-organized keyword system maintained by the National Library of Medicine.} serving essentially as the core concepts that classify the topic of a piece of published work. In this paper, we focused on the second-level MeSH terms under the ``Disease" branch (C-branch). This is because communicable and non-communicable diseases continue to be a major global health issue with significant impact on human society \cite{mcintoshGlobalFundingCancer2023,headAllocationUS105Billion2020}, and knowledge integration across diseases has made significant contribution to clinical and research knowledge \cite{duIntegrationVsSegregation2025a,schwetzExtendedImpactHuman2019a}. Note that we selected the second-level terms under the ``Disease" branch to ensure a good balance between granularity and computational feasibility of data gathering.

Following our prior work \cite{duIntegrationVsSegregation2025a}, 
we constructed co-occurrence networks based on all journal publications in 1999, 2010, and 2022 respectively, for research published in impactful (I) and non-impactful (NI) journals. Given rate and download limits on the PubMed API, selecting three point in time was chosen as a pragmatic approach to understand changes over time. Specifically, we started data collection for 1999, as this is the first year in which SJR medical rankings were made available. 2010 and 2022 were selected as a midpoint and endpoint that are both separated by a decade, allowing to capture the time-dependence of the networks.

Co-occurrence of two MeSH terms $i$ and $j$, $c_{ij}$, counts the number of journal articles that are indexed with both MeSH term $i$ and MeSH term $j$. Following \cite{duIntegrationVsSegregation2025a,eckHowNormalizeCooccurrence2009b}, the co-occurrences were then normalised through the cosine similarity  to reflect the extent of knowledge integration between $i$ and $j$, i.e., $w_{ij}=c_{ij}/\sqrt{c_{ii}c_{jj}}$, where $c_{ii}$ and $c_{jj}$ measure the number of articles indexed with term $i$ and $j$ respectively. A high value of $w_{ij}$ implies a strong correlation between the pair of MeSH terms, i.e., a pair of well-integrated concepts. 

Data collection and construction of the correlation network resulted in 296 nodes for all three years. Preliminary analysis showed that the resulting networks were not always connected. As we are interested in analysis of knowledge integration over the entire complex system of  medical research, we excluded MeSH terms that correspond to isolated nodes. Following this procedure, we obtain a core network composed of 201 nodes for all three years.



As preliminary analysis also indicated the possibility of sample size effects due to the much larger number of papers used to construct the NI network, we also constructed a version of the NI network based on less papers. This was done by restricting paper collection to only one month in the year when collecting NI data. After initial comparisons, which revealed no particular biases depending on the choice of month, without losing generality, we picked the month of June. 

\section{Result}\label{sec5}
\subsection{Medical Knowledge Network and Knowledge Clusters}
\begin{figure}[htp]
    \centering
    \includegraphics[width=\linewidth]{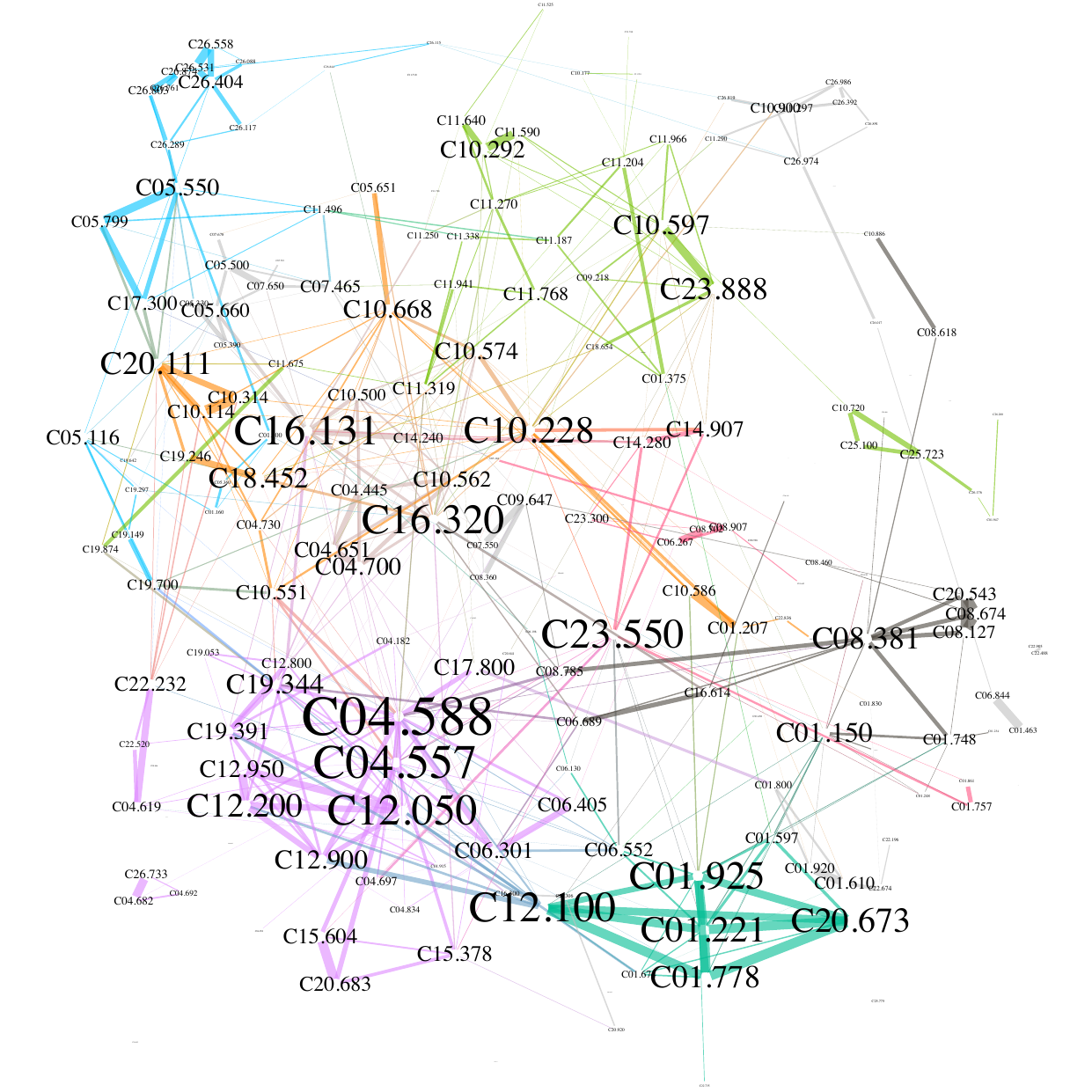}
    \caption{Impactful network in 1999. Links with strength less than $0.080$ were filtered out for better visualisation. Communities were detected based on the Louvain algorithm \cite{blondelFastUnfoldingCommunities2008}. We find a very similar community breakdown for the non-impactful network in 1999 (not shown).}
    \label{fig:1999I}
\end{figure}

To gain intuition, we start by visualising the impactful network in 1999 Fig \ref{fig:1999I}, where nodes are coloured by modularity decomposition using the Louvain method \cite{blondelFastUnfoldingCommunities2008}. We observe significant modularity ($Q=0.47$) and a breakdown into several major clusters, including a cancer-related cluster (C04 pink), an infectious diseases related cluster (C01 cold green), a nervous system diseases related cluster (C10 warm green), a respiratory diseases related cluster (C08 dark grey), and a musculoskeletal diseases related cluster (C05 blue). C16 (Congenital, Hereditary and Neonatal Diseases and Abnormalities) and C23 (Pathological Conditions, Signs and Symptoms) appear to be intermediaries that connect different clusters. 

\subsection{Sample Size effect}
\begin{figure}[htp]
    \centering
    \includegraphics[width=\linewidth]{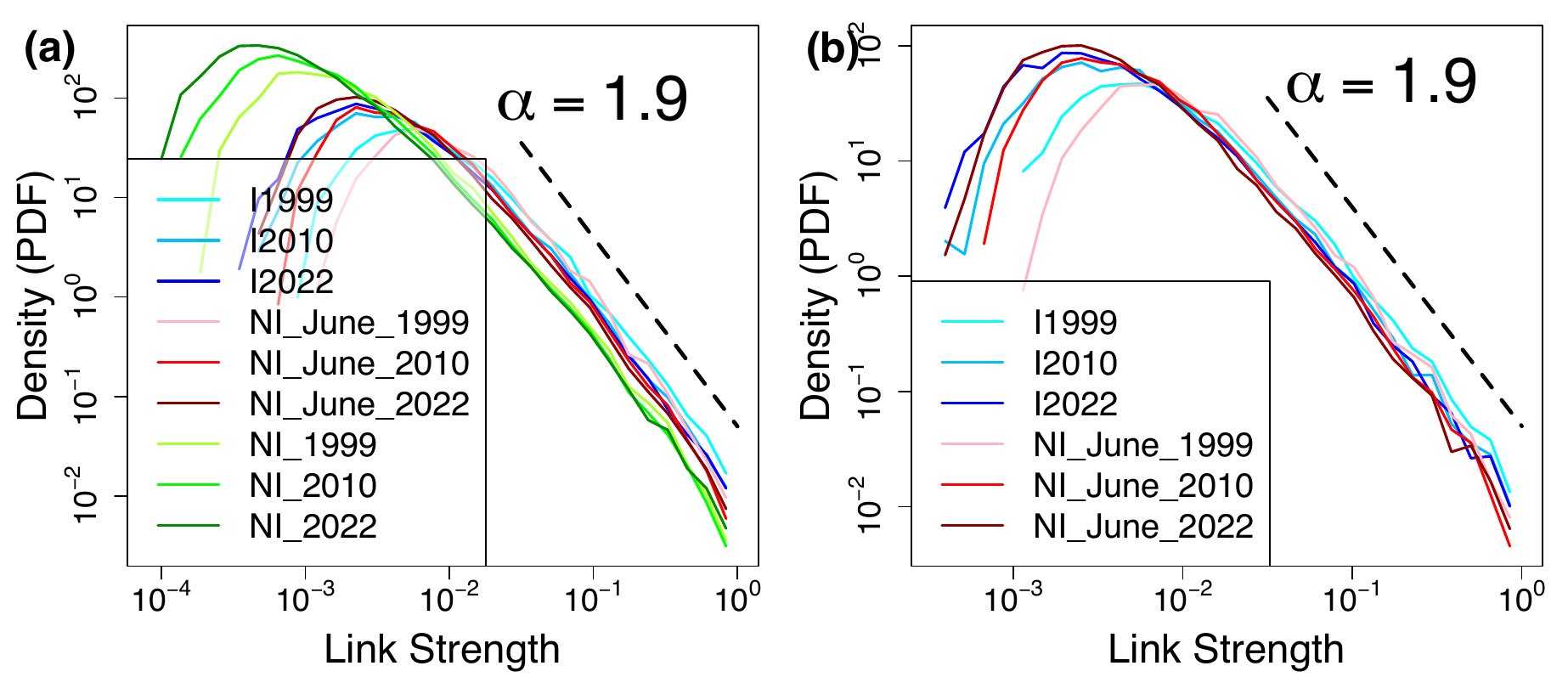}
    \caption{Link Strength Distribution for (a) I, NI, and NI-June networks, and (b) I and NI-June networks. Logarithmic binning was applied on x-axis with 30 bins. A power law distribution with exponent $\alpha=1.9$ and minimum value of link strength $x_{min}=10^{-1.5}$} is plotted as the dotted line against the tails of the distributions.
    \label{fig:link_strength}
\end{figure}

Plotting the link strength distribution for I and NI networks for 1999, 2010 and 2022 in Fig \ref{fig:link_strength}), we first note that the distributions show approximately power-law distributions over several orders of magnitude in the x and y dimensions with a power law exponent close to $2$. Next, we observe that NI networks tend to have a higher frequency of weaker links and less frequently stronger links than I (Fig \ref{fig:link_strength}.a). We hypothesize that this may be due to a sample size effect, as the network constructed for NI is based on a much larger set of papers than the I network Tab \ref{tab:paper number}. Adjusting for sample size, we note that the dominance of weak links disappears when comparing approximately equal samples (Fig \ref{fig:link_strength}.b, Tab \ref{tab:paper number}), i.e. the data for I and the NI sample collected for the month of June, which supports our hypothesis.


We further note that the existence of sample size differences chronologically in Tab \ref{tab:paper number}, where the number of papers has increased significantly across time for I, NI and NI-June. Thus, we leave the temporal analysis for future work and focus on the comparison between I and NI-June networks in this paper.

\begin{table}[htp]
    \centering
    \begin{tabular}{cccc}
    \hline
    Number of Papers  & 1999 & 2010 & 2022 \\
    \hline
    I & 13411 & 36507 & 56421\\
    \hline
    NI June & 16051 & 39971 & 68345 \\    
    \hline
    NI & 219827 & 404115 & 663182 \\   
    \hline
    \end{tabular}
    \caption{Total number of journal papers used to construct the networks. }
    \label{tab:paper number}
\end{table}

We plotted the node strength distribution for the I and NI-June networks. We note that the distributions shows approximately an exponential distribution for node strengths in the range between $1$ and $6$.
\begin{figure}[htp]
    \centering
    \includegraphics[width=0.5\linewidth]{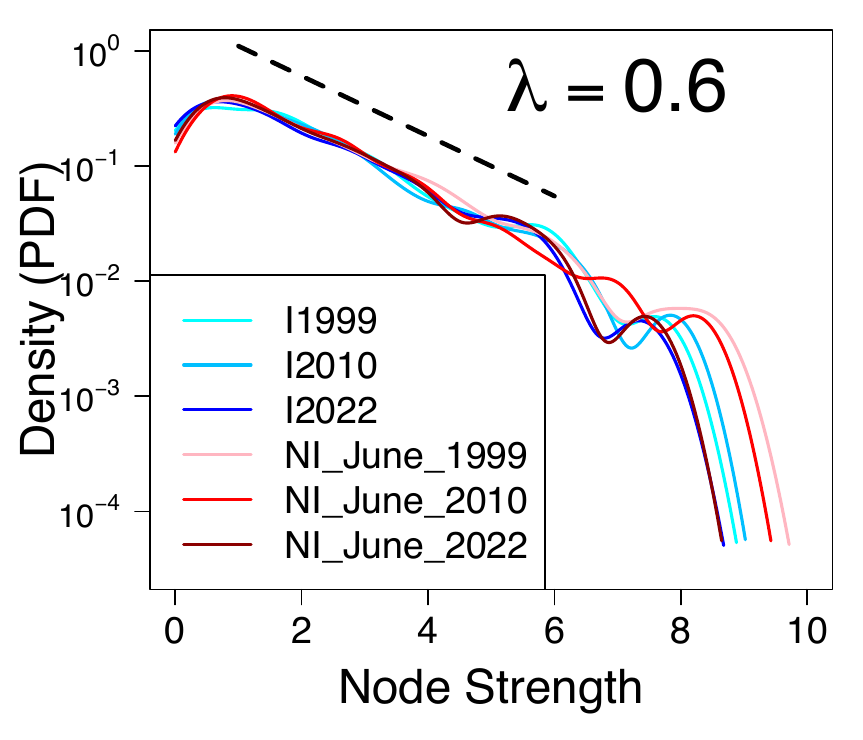}
    \caption{Node strength distribution of the I and NI networks. An exponential distribution with decay rate $\lambda=0.6$ with node strength $s\in[1,6]$ is plotted as the dotted line against the distributions.}
    \label{fig:degree}
\end{figure}

\subsection{Global Network Properties}
Comparing global network statistics in Tab \ref{tab: Global Net Measures}, we note that the NI-June network has a higher average node strength, or network coherence, $\overline{s}_{NI,June}$ than $\overline{s}_I$ of the I network. This indicates that non-impactful research is more interdisciplinary than impactful research. This result is also supported by the fact that the NI-June networks have consistently higher numbers of links $N_{NI,June}$ than $N_I$ of I, i.e., there are consistently more and stronger connections between medical concepts in non-impactful work than in impactful work.

We further note that the I network exhibits higher network modularity than the NI-June network (Tab \ref{tab: Global Net Measures}), despite that this effect becomes less pronounced in later years. This suggests that impactful interdisciplinary research tends to be more compartmentalised, i.e., focuses on smaller coherent knowledge clusters. Moreover, in terms of the global clustering coefficients and average shortest path lengths we find no significant differences between the I and the NI-June network.


\begin{table}[htp]
    \centering
    \begin{tabular}{cccc}
    \hline
     Global Network Measure  & 1999 & 2010 & 2022 \\
    \hline
     $\overline{s}_I$   & 1.75 & 1.71  & 1.64\\
     $\overline{s}_{NI,June}$ & 1.85 & 1.81 & 1.73\\
     
    \hline
     $N_I$   & 3689 & 4733 & 4898\\
     $N_{NI,June}$ & 4459 & 5926 & 6290\\
    \hline
    $Q_I$ & 0.47 & 0.45 & 0.45 \\
    $Q_{NI,June}$ & 0.41 & 0.41 & 0.42 \\
    \hline
    $GCC_I$  & 0.49 & 0.53 & 0.55 \\
    $GCC_{NI,June}$ & 0.50 & 0.55 & 0.58 \\
    \hline
     $ASPL_I$   & 26.28 & 26.52  & 29.54\\
     $ASPL_{NI,June}$ & 24.38 & 25.26 & 27.51\\
    \hline
    \end{tabular}
    \caption{Average node strength $\overline{s}$ \cite{rafolsDiversityNetworkCoherence2010}, number of links $N$, modularity $Q$ \cite{blondelFastUnfoldingCommunities2008}, global clustering coefficient $GCC$ \cite{onnelaIntensityCoherenceMotifs2005} and average shortest path length $ASPL$ \cite{brandesFasterAlgorithmBetweenness2001} for the I and NI-June networks. The shortest path length between any two node $i$ and $j$ is defined as $d(i,j)=min(\frac{1}{w_{ih}}+\cdots+\frac{1}{w_{hj}})$ with $h$ being intermediary nodes between node $i$ and $j$ \cite{brandesFasterAlgorithmBetweenness2001}.}
    \label{tab: Global Net Measures}
\end{table}

\subsection{Node Importance and Community-wide Knowledge Integration}
\label{sec:Node Importance and Community-wide Knowledge Integration}
Below, we continue by analysing node importance in the networks. The corresponding stats are reported in Tab \ref{tab:local measures}, where we focus on the individual ranking of nodes in various centrality measures. As we are mostly interested in node rankings regarding overall importance and bridging effects, we focus on measuring node strength $s_i$ and node betweenness $b_i$. We make the following observations. Neoplasm by site (C04.588) leads in node strength $s_i$ and node betweenness $b_i$ in both I and NI-June all the time, highlighting its crucial role in both integrating knowledge with relevant medical concepts and facilitating exchanges between distant knowledge clusters in medical research. We note that central Nervous System Diseases (C10.228) appears in the top two regarding betweenness centrality, but is not found in the top ranked nodes regarding strength. This indicates that research on CNS disease relatively lacks of local knowledge integration with other medical concepts but serves a strong bridge connecting rather distant bodies of knowledge. 

To gain further insights on the patterns of knowledge integration, we aggregated link strengths of each network at a higher level of aggregation in the MeSH hierarchy. For practical purposes and a high-level comparison, we chose the first level of category C in the MeSH hierarchy, as this only leaves us with 22 concepts. Each first-level term could be seen as a (pre-imposed) hierarchical community consisting of a number of second-level terms. We partitioned the resulting total community-wide connection strength $s_c$ into intra-community connection strength $s_c^{intra}$ and inter-community connection strength $s_c^{inter}$. Results for all three measures are reported in Tab \ref{tab:local measures}. Observing the results, it becomes clear that the ranking of $s_c$ is very stable over time, i.e., Infections (C01), Nervous System Diseases (C10) and Neoplasms (C04) ranks top three. 

High $s_c$ of Infections (C01) is dominant in intra-community connections (highest $s_c^{intra}$), while its role is less profound in the inter-community connection strength $s_c^{inter}$. This indicates that knowledge integration in infection-related medical research tends to reinforce existing linkages, which is confirmed by stronger connections between C01 terms in Fig \ref{fig:1999I}. On the other hand, Nervous System Diseases (C10) plays a much stronger role in inter-community connections by leading in the $s_c^{inter}$ ranking. This indicates that knowledge integration in nervous system diseases tends to bridge distant knowledge. This is confirmed in by observing the leading positions of C10 terms in top $b_i$ rankings in Tab \ref{tab:local measures}. 

Furthermore, we note Neoplasms (C04) manifests a relatively weaker intra-community connectivity (absence from top three $s_c^{intra}$) with a stronger inter-community connectivity (top two $s_c^{inter}$). Such a strong bridging role is again confirmed by the leading positions of C04 related terms in $b_i$ rankings.

\begin{table}[htp]
    \centering
    \begin{tabular}{llll}
    \hline
      \textbf{Top} $\mathbf{s_i}$  & 1999 & 2010 & 2022 \\
    \hline
       I  & \Centerstack{C04.588(7.62)\\C04.557(6.52) \\ C12.050(5.95)} & \Centerstack{C04.588(7.84)\\C04.557(6.48) \\ C23.550(6.23)} & \Centerstack{C04.588(7.35)\\C04.557(5.95) \\  C12.050(5.67)} \\
    \hline
       NI June  &\Centerstack{C04.588(8.41)\\C04.557(7.58) \\ C12.050(6.35)} &\Centerstack{C04.588(8.22) \\ C04.557(6.99)\\C23.550(6.79)} &\Centerstack{C04.588(7.43) \\ C12.050(5.95) \\C16.131(5.94)} \\
    \hline
      \textbf{Top} $\mathbf{b_i}$  &  &  &  \\
    \hline
       I  & \Centerstack{C04.588(0.23)\\C10.228(0.17) \\ C16.131(0.16)} & \Centerstack{C04.588(0.29)\\C10.228(0.21) \\ C16.131(0.13)} & \Centerstack{C04.588(0.25)\\C10.228(0.15) \\ C10.551(0.11)} \\
    \hline
       NI June  & \Centerstack{C04.588(0.25)\\C10.228(0.18) \\ C10.551(0.14)} & \Centerstack{C04.588(0.29)\\C10.228(0.17) \\ C10.551(0.12)} & \Centerstack{C04.588(0.24)\\C10.228(0.16) \\ C16.131(0.15)} \\
       \hline
      \textbf{Top} $\mathbf{s_c}$   &  &  &  \\
    \hline
       I   & \Centerstack{C01(43.74)\\C10(39.76)\\C04(33.86)} & \Centerstack{C01(41.61)\\C10(37.22)\\C04(31.65)} & \Centerstack{C01(38.34)\\C10(38.01)\\C04(28.16)} \\
    \hline
       NI June   & \Centerstack{C01(44.54)\\C10(42.06)\\C04(34.28)} & \Centerstack{C01(43.45)\\C10(39.51)\\C04(32.82)} & \Centerstack{C01(39.42)\\C10(39.35)\\C04(27.18)} \\
       \hline
      \textbf{Top} $\mathbf{s_c^{intra}}$   &  &  &  \\
    \hline
       I   & \Centerstack{C01(20.32)\\C26(13.97) \\C10(12.15)} & \Centerstack{C01(19.80)\\C11(12.40) \\C26(11.93)} & \Centerstack{C01(16.55)\\C10(12.77) \\C26(11.93)} \\
    \hline
       NI June  & \Centerstack{C01(19.85)\\C10(13.24) \\C26(11.40)} & \Centerstack{C01(19.43)\\C26(12.65) \\C10(12.27)} & \Centerstack{C01(16.33)\\C26(14.98) \\C10(12.46)} \\
       \hline
      \textbf{Top} $\mathbf{s_c^{inter}}$   &  &  &  \\
    \hline
       I  & \Centerstack{C10(27.61)\\C04(25.29) \\C01(23.41)} & \Centerstack{C10(25.44)\\C04(24.19) \\C01(21.81)} & \Centerstack{C10(25.24)\\C04(22.18) \\C01(21.78)} \\
    \hline
       NI June  & \Centerstack{C10(28.82)\\C04(26.94) \\C01(24.69)} & \Centerstack{C10(27.25)\\C04(25.82) \\C01(24.01)} & \Centerstack{C10(26.89)\\C01(23.09) \\C04(21.81)} \\
       
  \hline
    \end{tabular}
    \caption{Top three second-level MeSH terms, ranked by node strength $s_i$ and betweenness $b_i$. The total connection strength $s_c$ of each parent category (first-level MeSH term) is computed by aggregating intra-category strength $s_c^{intra}$ and inter-category strength $s_c^{inter}$. Top three first-level categories with respect to $s_c$, $s_c^{intra}$ and $s_c^{inter}$ are listed. C04.588: Neoplasms by Site. C04.557: Neoplasms by Histologic Type. C12.050: Female Urogenital Diseases and Pregnancy Complication. C10.228: Central Nervous System Diseases. C23.550: Pathologic Processes. C16.131: Congenital Abnormalities. C10.551: Nervous System Neoplasms. C01: Infections. C10: Nervous System Diseases. C04: Neoplasms. C26: Wounds and Injuries. C11: Eye Disease. }
    \label{tab:local measures}
\end{table}

\subsection{Analysing Difference between Impactful and Non-impactful Research}
\label{section:Impactful vs Non-impactful Research: A Difference Network}
To identify areas of research where work published in prestigious or less prestigious journals differs significantly, we explore how the I networks differ from the NI-June networks. To do this, we constructed difference networks with link strengths defined by the absolute difference between the link weights in the I and NI-June networks (Diff networks thereafter). 

\begin{figure}
    \centering
    \includegraphics[width=\linewidth]{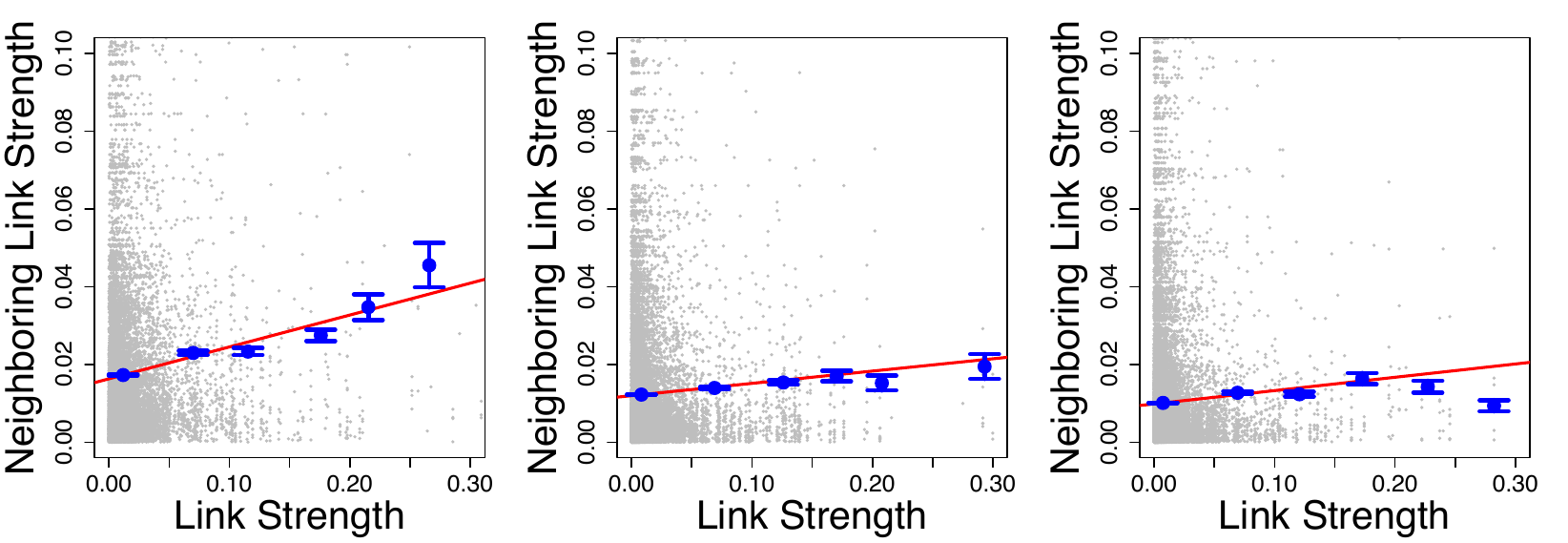}
    \caption{The relationship between link strength in Diff networks (x-axis) and neighbouring link strength (y-axis) in 1999 (left), 2010 (middle), and 2022 (right). The red line is a fitted linear regression line, with slopes of $0.08$ in 1999, $0.03$ in 2010 and $0.03$ in 2022, and $R^2$ of $0.007$ in 1999, $0.001$ in 2010 and $0.001$ in 2022. All three slopes reported are statistically significant ($p<2^{-16}$). Blue points represent the averages over bins, with error bars representing one standard error. } 
    \label{fig:assortativity}
\end{figure}

To explore the global patterns of connectivity of the Diff networks, we are particularly interested in the location of strong differences relative to each other. We first examine whether stronger links are more likely to be adjacent. To operationalise this idea, for the Diff networks, we consider the strength of all neighbouring links for each link and test whether stronger links tend to have stronger neighbouring links. Plotting link strength on the x-axis and neighbouring link strength on the y-axis in Fig \ref{fig:assortativity}, we note a small positive correlation, see the regression lines for each year. We conclude that there is a weak tendency for stronger links to pair with stronger neighbouring links in the Diff networks.

We further explore local patterns of connectivity of the Diff networks by examining node level differences. We first report the ranking of the node strength $s_i$ in Tab \ref{tab:diff_node}. We observe several nodes in the Neoplasms branch (C04) rank within the top five, including Neoplasms by Histologic Type, Neoplasms, Multiple Primary, and Neoplasms by Site. Besides, congenital Abnormalities ranks at top two in 1999 and 2022. Furthermore, pathologic Processes and Genital Diseases rank within the top five in all year. We conclude that differences between the networks tend to concentrate on areas related to cancer, congenital diseases, and pathological processes.

\begin{table}[htp]
\centering
\begin{tabular}{llll}
  \hline
\textbf{1999 C-LV2 Term}   & $s_i$ & C-LV1 Term & C-LV1 Code \\ 
  \hline
  Neoplasms by Histologic Type & 2.33 & Neoplasms & C04 \\ 
  Congenital Abnormalities & 2.23 & CHNDA & C16 \\ 
  Genital Diseases & 2.15 & Urogenital Diseases & C12 \\ 
  Neoplasms, Multiple Primary & 2.14 & Neoplasms & C04 \\ 
  Pathologic Processes & 2.12 & PCSS & C23 \\ 
    \hline  
  \textbf{2010 C-LV2 Term} &  &  &  \\ 
  \hline
  Pathologic Processes & 2.15 & PCSS & C23 \\ 
  Neoplasms by Histologic Type & 2.03 & Neoplasms & C04 \\ 
  Cranial Nerve Diseases & 1.90 & Nervous System Diseases & C10 \\ 
  Genital Diseases & 1.81 & Urogenital Diseases & C12 \\ 
  Neoplasms by Site & 1.80 & Neoplasms & C04 \\ 
  \hline
  \textbf{2022 C-LV2 Term }&  &  & \\ 
  \hline
Congenital Abnormalities & 1.97 & CHNDA & C16 \\ 
  Musculoskeletal Abnormalities & 1.88 & Musculoskeletal Diseases & C05 \\ 
  Neoplasms by Histologic Type & 1.71 & Neoplasms & C04 \\ 
  Stomatognathic System Abnormalities & 1.64 & Stomatognathic Diseases & C07 \\ 
  Pathologic Processes & 1.63 & PCSS & C23 \\ 
   \hline
\end{tabular}
\caption{Ranking of the node strengths $s_i$ of the Diff networks. The shorthands stand for: PCSS - Pathological Conditions, Signs and Symptoms, SCTD - Skin and Connective Tissue Diseases, CHNDA - Congenital, Hereditary, and Neonatal Diseases and Abnormalities.}
\label{tab:diff_node}
\end{table}

Next, we are interested in investigating differences in interdisciplinarity at the community level. For this purpose, we follow the procedure outlined in section \ref{sec:Node Importance and Community-wide Knowledge Integration} and report community-level connection strength $s_c$, intra- and inter-community connection strength, $s_c^{intra}$ and $s_c^{inter}$ for the Diff networks in Tab \ref{tab:diff_community}. We first note that Infections (C01), Eye Diseases (C11) and Wounds and Injuries (C26) rank at top three of $s_c$ in the Diff networks for all years. The same communities also lead in the $s_c^{intra}$ with slight reordering; moreover, Infections (C01) also leads in $s_c^{inter}$. These observations indicate that Infections (C01), Eye Diseases (C11) and Wounds and Injuries (C26) are the main areas where the patterns of knowledge integration differ significantly in terms of total community-level connectivity, intra- and inter-community connectivity. 

We also note that Neoplasms (C04), Musculoskeletal Diseases (C05) and Nervous System Diseases (C10) rank top two or three in the $s_c^{inter}$ ranking yet are absent from top ranked communities in $s_c^{intra}$. We conclude that I networks differs from NI networks significantly in terms of the inter-community connectivity associated with these areas, and less significantly in terms of the intra-community connectivity.
\begin{table}[htp]
    \centering
    \begin{tabular}{lccc}
    \hline
      \textbf{1999}   & $s_c$ & $s_c^{intra}$  & $s_c^{inter}$\\
      \hline
        &C01(24.46) &C26(12.90) &C01(15.32) \\
        &C26(23.60) &C11(9.59) &C10(15.18) \\
        &C11(21.46) &C01(9.13) &C04(12.30) \\
    \hline
      \textbf{2010}   & & & \\
      \hline
        &C01(22.15) &C01(8.50) &C01(13.65) \\
        &C11(18.09) &C26(8.32) &C10(11.52) \\
        &C26(16.89) &C11(7.94) &C04(10.37) \\
    \hline
      \textbf{2022}   & & & \\
      \hline
        &C01(16.14) &C26(7.90) &C01(10.76) \\
        &C11(15.80) &C11(7.42) &C10(10.07) \\
        &C26(15.16) &C01(5.37) &C05(9.19) \\
    \hline
    \end{tabular}
    \caption{Ranking of total community-level connection strength $s_c$, intra-community connection strength $s_c^{intra}$ and inter-community connection strength $s_c^{inter}$ of Diff networks. The largest differences are observed for C01: Infections. C10: Nervous System Diseases. C04: Neoplasms. C05: Musculoskeletal Diseases. C26: Wounds and Injuries. C11: Eye Disease.}
    \label{tab:diff_community}
\end{table}

\section{Conclusion}
In this paper, we constructed and compared networks of interdisciplinary knowledge structure for research published in prestigious and less-prestigious medical journals. We draw two main conclusions to answer the two questions we started with. First, to answer our first question regarding whether prestigious journals produce more interdisciplinary knowledge than less prestigious journals in medicine,
we found research from prestigious journals tends to be less interdisciplinary than research from other journals. We also established that, as it bridges distant knowledge clusters, cancer-related research can be seen as the main driver of interdisciplinarity in medical science. 

Second, as concerns our second question related to in which research areas do prestigious and less prestigious medical journal differ in terms of interdisciplinary knowledge production, we constructed the absolute difference network for research from prestigious and less prestigious journals and found a weak tendency for differences in topic correlations to be co-located. Then, we identified broad research topics like infections with significant intra-topic difference in interdisciplinarity, and topics like nervous system diseases and cancer with significant inter-topic difference in interdisciplinarity


In conclusion, our paper provides a novel comparison of interdisciplinarity of publications in more and less prestigious journal in medicine. Interdisciplinary research is the key to bring innovation to address complex problems. Interdisciplinarity in medical science could benefit from prestigious journals easing rigid disciplinary boundaries.



%
%

\end{document}